\documentstyle[11pt]{article}
\begin{document}

\begin{center}
{\bf 
Generalized Short Pulse Equation \\
for Propagation of Few-Cycle Pulses in Metamaterials}

Monika E. Pietrzyk and Igor V. Kanatt\v sikov

School of Physics and Astronomy, University of St Andrews\\
North Haugh, KY16 9SS St Andrews, UK\\
mp212@st-andrews.ac.uk

{\bf Abstract:} 
We show that propagation of ultrashort (few-cycle) pulses in nonlinear
Drude metamaterials with both electric and magnetic Kerr nonlinearities is
described by coupled generalized Short Pulse Equations. The resulting system of
equations generalizes to the case of metamaterials both the Short Pulse 
Equation and its vector generalizations. 



\end{center}

Metamaterials are artificial structures that display properties beyond those 
available in naturally occurring materials. The most notable are the negative 
refraction index materials with simultaneously negative electric and magnetic 
dispersive responses, which affect substantially the conventional optics and 
its applications. Metamaterials with the negative refraction index have a 
number of extraordinary properties, such as the reversed Snell refraction, 
reversed Doppler effect, reversed radiation tension, negative Cerenkov 
radiation, reversed Goos-H\"anchen shift, etc. Such materials can be used in 
various devices, such a compact cavities, superlenses, subwavelength
waveguides and antennas, electromagnetic cloaking devices, tunable mirrors,
isolators, phase compensators, and many others. Metamaterials can be also 
used to study ideas developed to describe physics in curved space-times and 
to model virtually any space-time metric of General Relativity.

A typical metamaterial with the negative refraction index is composed of a
combination of a regular array of electrically small resonant particles, 
referred to as split-ring resonators, and a regular array of conducting wires 
responsible, respectively, for the negative electric permittivity and 
negative magnetic 
permeability. The size and spacing of these elements is supposed to be much 
smaller than the wavelength of the propagating optical field, so that the 
metamaterial can be considered as a homogeneous medium.

Different models have been used to describe propagation of short and ultrashort
optical pulses in metamaterials. For metamaterials without nonlinear 
magnetization a generalized nonlinear Schr\"o-dinger equation (NSE) has been 
derived [1, 2, 3, 4, 5]. For metamaterials with nonlinear magnetizations a 
single-component NSE for the electric field has been obtained in [6]. For the 
sufficiently long temporally optical pulses a system of coupled NSE can be 
used [7, 8, 9, 10]. 

However, for ultrashort few-cycle pulses the envelope approximation is not valid
and the NSE cannot be applied. For metamaterials without the nonlinear
magnetization a single-component generalized SPE for the electric field has been
obtained in [11, 12]. When the nonlinear magnetization is present the 
equations for the electric and magnetic field cannot be uncoupled [13].
Here we show that the propagation of few-cycle pulses in metamaterials with 
electric and magnetic Kerr-type nonlinear response can be described by a 
coupled system of Short Pulse Equations for the electric and magnetic field.

It has been shown that a composite metamaterial with negative refraction index 
can develop a non-linear macroscopic magnetic response. This means that, 
although the host medium has a negligible magnetic nonlinearity, the periodic 
inclusions of the metamaterial can produce an effective magnetic nonlinear 
response.

Thus, let us consider a metamaterial with nonlinear magnetization. Let us 
start from the Maxwell equations for an optical field propagation along the 
z-direction.
\[ \partial_z E = - \partial_t B - \partial_t M_{nl}, \]
\[ - \partial_z H = \partial_t D + \partial_t P_{nl}, \]
\[ \partial_x D =0, \]
\[ \partial_y B =0,\]
where it is assumed that the electric and magnetic fields are linearly 
polarized:
\[ {\bf E}= (E, 0, 0), \hspace{5mm} 
{\bf H} = (0, H, 0).\]
The dielectric and magnetic response of a nonlinear material is 
characterized by electric displacement field 
$\tilde{D} (\omega) = \epsilon(\omega) \tilde{E}(\omega)$,
magnetic induction $\tilde{B}(\omega) = \mu(\omega) \tilde{H}(\omega)$,
nonlinear polarization $P_{nl}= \epsilon_{nl} E$, and nonlinear magnetization
$M_{nl} = \mu_{nl} H$.

Let us assume that both the electric and the magnetic nonlinearities are of 
Kerr type, $\epsilon_{nl} = \chi_e E^2$ and $\mu_{nl} = \chi_m H^2$. 
Substituting the material equations into the Maxwell equations we obtain in 
the frequency domain:
\begin{equation}
\partial_z \tilde{E} = -i \omega \mu (\omega) \tilde{H} 
- i \omega \chi_m {\tilde H}^3,
\label{maxe}
\end{equation}
\begin{equation}
\partial_z \tilde{H} = -i \omega \epsilon(\omega) \tilde{E} 
- \omega i \chi_e {\tilde E}^3.
\label{maxh}
\end{equation}
Acting with $\partial_z$ on equation (\ref{maxe}) and using equation 
(\ref{maxh}) we get:
\begin{equation}
\partial_{zz} \tilde{E} = -\omega^2 \left( \mu(\omega) \epsilon(\omega)
+ \mu(\omega) \chi_e \tilde{E}^2 + 3 \epsilon(\omega) \chi_m \tilde{H}^2
+ 3 \chi_e \chi_m \tilde{E}^2 \tilde{H}^2 
\right) \tilde{E}. 
\label{fielde}
\end{equation}
Similarly, acting with $\partial_z$ on equation (\ref{maxh}) and using equation 
(\ref{maxe}) we obtain:
\begin{equation}
\partial_{zz} \tilde{H} = -\omega^2 \left( \epsilon(\omega) \mu(\omega)
+ \epsilon(\omega) \chi_m \tilde{H}^2 + 3 \mu(\omega) \chi_e \tilde{E}^2
+ 3 \chi_e \chi_m \tilde{E}^2 \tilde{H}^2 
\right) \tilde{H}.
\label{fieldh} 
\end{equation}

Now, let us assume that the dispersive properties of the metamaterial are 
given by the lossless Drude model:
\[\epsilon(\omega) = \epsilon_0 ( 1 - \omega_e^2/\omega^2 ) 
\hspace{5mm} \mbox{and} \hspace{5mm}
\mu(\omega) = \mu_0 ( 1 - \omega_m^2/\omega^2 ),\]
where $\omega_e$ and $\omega_m$ are the electric and magnetic plasma 
frequencies, respectively. Then
\[ \epsilon(\omega) \mu(\omega) \approx \epsilon_0 \mu_0 
\left(1 - \frac{\omega_e^2}{\omega^2} - \frac{\omega_m^2}{\omega^2}
+\frac{\omega_e^2 \omega_m^2}{\omega^4} \right). \]
Neglecting the term proportional to $\omega^{-4}$ and using 
$\epsilon_0 \mu_0 = 1/c^2$, where $c$ is the velocity of light in vacuum, 
we obtain:
\[\epsilon(\omega) \mu(\omega) \approx 
\epsilon_0 \mu_0 (1 - \omega_e^2/\omega^2 - \omega_m^2/\omega^2). \]

Substituting these formulas into equations (\ref{fielde}) and (\ref{fieldh}),
applying the Fourier transform and neglecting higher order nonlinear terms 
proportional to $E^2 H^2$ we obtain:
\begin{equation}
\partial_{zz} E = \frac{\partial_{tt} E}{c^2} 
+ \frac{\omega_e^2 + \omega_m^2}{c^2} E
+\mu_0 \chi_e \partial_{tt}(E^3)
+\mu_0 \omega_m^2 \chi_e E^3 
+ 3 \epsilon_0 \chi_m \partial_{tt} (H^2 E)
+ 3 \epsilon_0 \omega_e^2 \chi_m H^2 E
\label{eqe}
\end{equation}
and
\begin{equation}
\partial_{zz} H = \frac{\partial_{tt} H}{c^2} 
+ \frac{\omega_e^2 + \omega_m^2}{c^2} H
+\epsilon_0 \chi_m \partial_{tt}(H^3)
+\epsilon_0 \omega_e^2 \chi_m H^3 
+ 3 \mu_0 \chi_e \partial_{tt} (E^2 H)
+ 3 \mu_0 \omega_m^2 \chi_e E^2 H.
\label{eqh}
\end{equation}

Introducing new variables: $\tau = t - z/c$ and $\zeta = z$ for which 
$\partial_{zz} = 1/c^2 \partial_{\tau \tau} + 2/c \, \partial_{\zeta \tau} +
\partial_{\zeta \zeta}$ and $\partial_{tt} = \partial_{\tau \tau}$ in equations 
(\ref{eqe}) and (\ref{eqh}) and making use of the paraxial
approximation: $\partial_{\zeta \zeta} E = \partial_{\zeta \zeta} H =0$, we obtain:
\begin{equation}
\partial_{\zeta \tau} E = \frac{\omega_e^2 + \omega_m^2}{2c} E
+ \frac 12 \mu_0 \chi_e c \left( \partial_{\tau \tau} + \omega_m^2 \right) E^3 
+ \frac 32 \epsilon_0 \chi_m c \, (\partial_{\tau \tau} + \omega_e^2) H^2 E
\label{spes}
\end{equation}
and
\begin{equation}
\partial_{\zeta \tau} H = \frac{\omega_e^2 + \omega_m^2}{2c} H
+ \frac 12 \epsilon_0 \chi_m c \left(\partial_{\tau \tau} + \omega_e^2\right) H^3 
+ \frac 32 \mu_0 \chi_e c \, (\partial_{\tau \tau} + \omega_m^2) E^2 H.
\label{speh}
\end{equation}

Thus, we have obtained the set of two coupled generalized Short Pulse 
Equations. In the limit $\omega_m \rightarrow 0$ and $\chi_m \rightarrow 0$ 
equation (\ref{spes}) reduces to the Short Pulse Equation derived by 
Sch\"afer and Wayne [16]
\begin{equation}
\partial_{\zeta \tau} E = \frac{\omega_e^2}{2c} E
+ \frac 12 \mu_0 \chi_e c \, \partial_{\tau \tau}(E^3),
\label{spee}
\end{equation}
which later was shown to be an integrable system [15,16]. With certain
combination of the parameters of the metamaterial we can also obtain from
equations (\ref{spee}) and (\ref{speh}) different integrable vector 
generalizations of the short pulse equation obtained by us [17] and other 
authors [18,19].

In conclusion, a consideration of the propagation of ultra-short few-cycle 
polarized optical pulses in the Drude metamaterial optical fibers with 
electric and magnetic Kerr nonlineratity leads to the coupled set of equations 
which generalizes the Short Pulse Equation. It allows to describe the 
ultra-short and spectrally broad optical pulses beyond the slow varying 
envelope approximation. It may open a possibility of studying a new class of 
optical phenomena in metamaterials when the spectral range of the optical 
field overlaps the regions with different signs of optical indices of the 
metamaterial.

REFERENCES.

[1] M. Scalora, et al., Phys. Rev. Lett. 95, 013902 (2005),

[2] S. Wen, et al., Opt. Expres 14, 1568 (2006),

[3] S. Wen, et al., Phys. Rev. E 73, 036617 (2006),

[4] Hu Yong-Hua, et al., Chin. Phys. 15, 2970 (2006),

[5] A. Kumar Mishra and A. Kumar, J. Mod. Opt. 59, 1599 (2012),

[6] J. Zhang, el al., Phys Rev. A 81, 023829 (2010),

[7] I. Kοurakis and P. K. Shukla, Phys. Rev. E 72, 016626 (2005),

[8] N. Lazarides and G. P. Tsironis, Phys. Rev. E 71, 036614 (2005)

[9] S. Wen, et al., Phys. Rev. A 75, 033815 (2007),

[10] N. L. Tsitses, et al., , Phys. Rev. E 79, 037601 (2009),

[11] N. L. Tsitsas, et al., Phys. Lett. A 374, 1384 (2010),

[12] Y. Shen, et al., Phys. Rev. A 86, 023841 (2012),

[13] P. Kinsler, Phys. Rev. A 81, 023808 (2010),

[14] T. Sch\"afer and C.E. Wayne, Physica D 196, 90 (2004),

[15] A. Sakovich and S. Sakovich, J. Phys. Soc. Japan 74, 239 (2005),

[16] J. C. Brunelli, J. Math. Phys. 46, 123507 (2005).

[17] M. E. Pietrzyk, I. Kanattsikov, et al., J. Nonl. Math. Phys. 15, 
162 (2008).

[18] J.C. Brunelli, S. Sakovich, Phys. Lett. A 377, 80 (2012),
J. Math. Phys. 54, 012701 (2013).

[19] Y. Matsuno, J. Math. Phys. 52, 123702 (2011).

\end{document}